\documentstyle[12pt,fullpage]{article}
\newtheorem{defn}{Definition}

\newtheorem{prop}{Proposition}

\newcommand{\rarr}{\rightarrow}
\newcommand{\smaran}{{\sf Smaran}}
\newcommand{\tgr}{{\sf TGR}}

\begin{document}
\thispagestyle{empty}
\newtheorem{example}{Example}
\newtheorem{progs}{Program}
\begin{center}
{\Large \bf Static Analysis Techniques for Equational Logic
Programs\footnote{Partially supported by NSF grant CCR-9732186}}  
\vspace{2mm}  

{\large Rakesh Verma} \vspace{1mm}

Computer Science Department \\
University of Houston \\
Houston, TX 77204 \\
Ph: (713)743-3348 \\
Fax: (713)743-3335 \\
Email: rmverma@cs.uh.edu \\
\end{center}
\begin{center}

{\bf Abstract}
\end{center}
An equational logic program is a set of directed equations or rules,
which are used to compute in the obvious way (by replacing equals with
``simpler'' equals). We present static analysis techniques for efficient
equational logic programming, some of which have been implemented 
in  $LR^2$, a laboratory for developing and evaluating fast,
efficient, and practical rewriting techniques.  Two novel
features of $LR^2$ are that non-left-linear rules are allowed in most
contexts and it has a tabling option based on the congruence-closure based
algorithm to compute normal forms. Although, the focus of this
research is on the tabling approach some of the techniques are
applicable to the untabled approach as well. Our presentation 
is in the context of $LR^2$, which is an interpreter, but some of the 
techniques apply to compilation as well.

\newpage
\pagenumbering{arabic}

\section{Introduction}
Equational logic programming has received increasing attention in
recent years and there are now fast compilers such as Elan, Maude,
etc., and fast interpreters such as $LR^2$. The objective of this
paper is to enhance the efficiency of interpretation (and possibly 
also compilation) of equational logic programming by analyzing the input
program. Since we want fast and efficient interpreters, our focus is on
practical and effective techniques that can be implemented very
cheaply (linear or close to linear time) in terms of their time and
space cost.  
To describe the techniques we assume the usual scenario, viz., an
equational program, or a set of rules,  $P$ and a term $t$ are given
and the goal is to compute a normal form of $t$. We  assume here
that $P$ is confluent and that $t$ is well-formed but we do not assume termination. Although most
of the techniques we present are of general significance, their immediate
motivation is to enhance the efficiency of $LR^2$, especially its
tabling mechanism, so we give a brief
self-contained introduction to $LR^2$  and  a description of
the tabling algorithm. For more details on $LR^2$, including useful optimizations and some performance results,  the reader can consult
\cite{rakshal}. Throughout this paper, when we use the term tabling we
are specifically referring to this algorithm. 
\subsection{$LR^2$}
$LR^2$ is a laboratory for developing and evaluating fast, efficient, 
and practical rewriting techniques. 
	$LR^2$ consists of an expression graph interpreter \tgr,
and an expression graph rewriter that tables the history of its reductions,
called \smaran\ , based on the  congruence closure approach. The input to $LR^2$ is a program
representing either an orthogonal, not necessarily terminating,
rewrite system, or a convergent (terminating and confluent) rewrite
system,  and an input   expression. First-order orthogonal systems can
easily express first-order functions and it is a well-known fact that
parallel-outermost reduction or lazy evaluation is necessary and
sufficient to compute with such systems. Observe that since convergent
systems can be given to $LR^2$, hence, it allows non-left-linear rules
such as $x + -x \rarr 0$ and $equal(x,x) \rarr true$, etc.  Similar to
algebraic  specification languages like OBJ, ASF+SDF and Elan a
program is composed from modules. Each module defines its own
signature and rewriting rules.  A module can import other
modules. Expressions in $LR^2$ are written in prefix form. The
language of $LR^2$ contains several built-in datatypes, viz.,
integers, floating-point arithmetic, booleans, characters, sets,
multisets, and strings with associated operations. The set datatype
supports the insertion, deletion, membership, union, etc., operations.
The string datatype supports membership and indexing operations.  

$LR^2$ also includes a variant detector that can determine if a new expression
is an alphabetic variant of an existing expression, which is currently usable with 
the tabling option. If so, the appropriate
variant of the result computed for the existing expression is used for
further rewriting instead of starting from scratch. 
$LR^2$ also allows a  compact form for storing lists of arithmetic
progressions as these occur frequently. Instead of storing the
entire list, $LR^2$ stores the initial value, the final value and the
difference.  $LR^2$ provides a set of commands so that it can  be called by other systems for symbolic computation. This currently requires the UNIX message  
passing mechanism.

   $LR^2$ provides a variety of options for controlling
the amount of history that is stored by the system, if 
the user chooses the tabling option. The default option using \smaran\
is to save the results of each rewrite 
step in a compact data structure. The language of $LR^2$ allows annotating  specific functions with the keyword ``memo''. This  
allows to save all the results of  rewriting expressions that have the
specified function symbol at the root. The Delete (history) option in
$LR^2$ allows to delete the entire 
history of rewrites performed so far except for the given expression and its
latest reduct after every  $i^{th}$   rewrite step, where  $i$  can be
specified by the user. Another possibility allowed is to delete the
entire history except the given expression and the latest reduct as soon as
the free memory available to the user drops to a user-specified
percentage of the total available memory.     Options can be combined in any way to suit the application. However, the delete option overrides the
other options.

Operators can also be declared associative (A) or commutative (C) or
both (AC) in $LR^2$. However, currently only
left-linear rules with A, C, or AC operators can be handled; efficient matching
algorithms/heuristics for nonlinear rules with such operators are
fairly involved and is currently in the testing
phase. It is known that A-matching, C-matching and AC-matching of
general patterns are all NP-complete even with some 
restrictions on occurrences of variables \cite{rakinfcom}. 

The rest of this paper is organized as follows. In Section 2, we
give a description of the main algorithm for \smaran\ ,  
in Section 3 we discuss techniques to enhance the efficiency of 
$LR^2$,  in Section 4 we
mention the techniques that have already been implemented in $LR^2$. We
conclude with some promising directions for future research. 
We omit here all proofs of the results stated in this paper. 

\section{Smaran's Core Algorithm } \label{ccnaalg}
The basic algorithm at the core of  \smaran\ was proposed in \cite{chewalg}
for orthogonal rules and later extended in \cite{rakdiss,rakfocs} to non-left-linear
rewrite systems under fairly general conditions. The basic algorithm
in turn is an extension of the 
 well-known congruence closure algorithm (CCA) \cite{dst,no,shostak,kozen}) for ground equations.  

Recall that CCA divides the set of terms
into numbered equivalence classes. Membership
of a term in an equivalence class is decided by its {\em
signature}. The signature of a term $f(t_1,\ldots , 
t_n)$ is the tuple  $\langle f~ \#[t_1]~\ldots \#[t_n]
\rangle$, where $\#[t_i]$ is the number of the equivalence class
containing the signature representing $t_i$. Equivalence class 
$C$ {\em represents} term $t$ if $C$ contains a signature representing
$t$. CCA operates by merging equivalence class 
representing terms whose equivalence follows from the given equations.

To extend CCA for normalization we make use of the concept of 
a distinguished signature in each equivalence class called the {\it
unreduced signature} \cite{chewalg,rakdiss}. Using this signature we
construct a distinguished term. It has been shown in
\cite{chewalg,rakdiss,rakfocs} that it is enough to examine this  
term to select useful rule instances, and that it is sufficient
to check this term for irreducibility for rewrite systems satisfying
fairly general conditions. If it exists and is irreducible, then the class 
containing this term has a normal form. 
Thus, instances of left hand sides  represented by reduced
signatures, do not lead to any progress towards the  normal form, and
any term represented by a reduced signature cannot be in normal
form. 
Whenever a rule instance $A \rarr B$ can be applied to
the distinguished term of a class, $C$, because of a match, the
signature  representing $A$ is marked reduced and the class 
representing $B$ (if any) is merged with  $C$. If there is no
class representing $B$, then a signature  representing $B$ is
constructed, inserted into $C$, and marked the
unreduced signature of $C$. 

The algorithm starts by constructing the signature of the given term. This signature is then 
inserted into a class and marked the unreduced signature of the class.
This class number is tracked throughout the process of normalization.
Signatures of terms are constructed in the obvious bottom-up way. 
We illustrate this algorithm with a
small example. Consider the  convergent rewrite system with three rules that are numbered for convenience, 
\[
1: fib(x)  \rarr f(x   > 1, x), 2: f(true,x)  \rarr  fib(x-1) +
fib(x-2), 3: f(false,x)  \rarr  1, \] that
uses built-in arithmetic to define the fibonacci function and 
let the input term be $fib(2)$. 

The initial set of classes is: \[ 0:\{2^{*}\}\hspace{0.05in} 1:\{\langle fib~
0 \rangle^{*}\} \] (the symbol `*' indicates unreduced
signatures). Now the match procedure  is called to find a match  
between the unreduced signature of any class and the left-hand side of
any rule. A match occurs between  class 1 and rule 1.
Now, the signature representing  the corresponding instance of the
right-hand side of rule 1   is constructed, inserted into class 1 and
marked as its unreduced signature. Note that here  we do
not show the signatures, corresponding to the built-in datatypes, that
can be evaluated directly.  At this stage the classes are as follows: 

\centerline{ $0:\{2^{*}\}\hspace{0.05in} 1:\{\langle fib~ 0 \rangle, \langle f~ 3~ 0 \rangle^{*}\}
\hspace{0.05in}2:\{1^{*}\}\hspace{0.05in}3:\{true^* \}$}

During the next iteration class 1 matches rule 2. The right-hand side
instance is $fib(1) + fib(0)$. The signature representing this is
created and inserted into class 1 as its new unreduced signature. At
the end of the second iteration the new/changed classes 
are as follows: 

\centerline{$ 1:\{\langle fib~ 0 \rangle, \langle f~ 3~ 0 \rangle, \langle +~ 4~
6 \rangle^{*}\}\hspace{0.05in}4:\{\langle fib~ 2 \rangle^*\} \hspace{0.05in}5:\{0^*\}
\hspace{0.05in} 6:\{\langle fib~ 5 \rangle^*\}$}

After two rewrite steps, using rules 1 and 3 respectively, the
term $fib(1)$ represented by class 4 reduces to 1, which is represented by
class 2. Hence classes 4 and 2 are merged, say into 2, and we have:

\centerline{ $0:\{2^{*}\}\hspace{0.02in} 1:\{\langle fib~ 0 \rangle, \langle f~ 3~ 0 \rangle, \langle +~ 2~ 6 \rangle^{*}\}
\hspace{0.02in}2:\{\langle fib~ 2 \rangle, \langle f~ 7~ 2 \rangle,
1^{*}\} $ }
\centerline{$3:\{true^* \} \hspace{0.02in}5:\{0^*\}
\hspace{0.02in} 6:\{\langle fib~ 5 \rangle^*\} \hspace{0.02in} 7:\{false^*\}$}

Note that signatures containing the class number 4 have been
updated to contain class number 2. After two more rewrite steps, the
term $fib(0)$ represented by class 6 reduces to 1. Hence classes 6
and 2 are merged, say into 2, and we get:

\centerline{$0:\{2^{*}\}\hspace{0.02in} 1:\{\langle fib~ 0 \rangle, \langle f~ 3~ 0 \rangle, \langle +~ 2~ 2 \rangle^{*}\}
\hspace{0.02in}2:\{\langle fib~ 2 \rangle, \langle fib~ 5 \rangle, \langle f~ 7~ 5 \rangle, \langle f~ 7~ 2 \rangle,
1^{*}\}$} 
\centerline{$3:\{true^* \} \hspace{0.02in}5:\{0^*\}  \hspace{0.02in}
 \hspace{0.02in} 7:\{false^*\}$}

Now, the unreduced signature of class 1 can be evaluated to yield the
term 2, which is in class 0. Hence, classes 1 and 0 are merged, say
into 0, and we get:

\centerline{$1:\{2^{*}, \langle fib~ 0 \rangle, \langle f~ 3~ 0 \rangle, \langle +~ 2~ 2 \rangle\}
\hspace{0.02in}2:\{\langle fib~ 2 \rangle, \langle fib~ 5 \rangle, \langle f~ 7~ 5 \rangle, \langle f~ 7~ 2 \rangle,
1^{*}\}$}
\centerline{$3:\{true^* \} \hspace{0.02in}5:\{0^*\}  \hspace{0.02in}
 \hspace{0.02in} 7:\{false^*\}$}

No more matches are found. Therefore, the algorithm checks for the
existence of a normal form for the given term. Class 1 contains the unreduced
signature {\it 2}, which is irreducible. Thus the normal
form of $fib(2)$ is 2.  Note that if the normal form 
of $fib(fib(2))$ is needed, no more computations are needed 
since this term is also represented by the signature $\langle fib~
0 \rangle$ in class 1. It is  simply extracted from class 1.  The
normal form of this term is also $2$. On the other hand, an
interpreter that does not store history would compute $fib(2)$ twice
to normalize this term.
This  compact data structure helps exploit the advantages of
storing history and can also speed up  normalization. 

\section{When does tabling help?}
In this section, we present techniques that analyze the program $P$ and
term $t$ to determine whether tabling could  be helpful in  computing
the normal form of $t$. Recall that tabling can help in two important
ways: 
\begin{enumerate}
\item It can improve the termination characteristics, i.e., tabled
rewriting can halt in cases where untabled would fail to halt because
there is no normal form. For example, consider the rule $a \rarr
f(a)$. 
\item It can improve the efficiency of the computation. Each rule
instance is applied at most once and its results  are stored. This is
extremely  useful for: (a) problems that recursively solve lots of
overlapping subproblems, e.g., dynamic programming problems, (b)
proving theorems of the form $A = B$, and determining whether a
critical-pair ($A, B$) generated during KB-completion \cite{kb} is
trivial. (It is possible that $A$ and $B$ were reduced to a common
term albeit at different times, since  this common term is remembered
in tabling, it will terminate immediately whereas untabled methods may
go on forever.), and (c) incremental computation. 
\end{enumerate}

We now develop two sets of necessary conditions that a program $P$
must satisfy so that tabling can potentially be useful for $P$, one
set for each way in which tabling can help.
\subsection{When does tabling improve termination?}
First we give necessary conditions on $P$ alone for tabling to help
termination. Next, we will analyze both $P$ and $t$ to develop even
stronger necessary conditions. 
As usual, we partition the constants and function symbols of $P$ into two classes:
constructors and defined symbols. 
\begin{defn}
A symbol is called {\em defined}
if it is the outermost symbol of the left-hand side of some rule in
$P$,
otherwise it is a constructor. 
\end{defn}
For example, $fib$ is a defined symbol in the program given above,
whereas $true$ is a constructor. 
Note that constructors include the constants from the predefined types
such as bool, float, int, char, etc., (we know that there can be no rules with integer, float, char and
bool constants as left-hand sides). 
The standard operators over these predefined 
 types  are treated as constructor functions in this section.
\begin{defn}
Let $f$ be a defined symbol of $P$  and let $g$ be any function symbol
or constant of $P$. We say
that $f$ {\em needs} $g$ if $g$ appears {\em in} the right-hand side of any
rule that has $f$ as its defined symbol.
\end{defn}

Now we define the {\em needs} graph of $P$ as the directed graph $G= (V,E)$,
where $V$ contains a vertex for  each function symbol or constant of
$P$ and $E$ contains a directed edge from the vertex for  $f$ to the
vertex for $g$ precisely when $f$ needs $g$.  Note that this graph may
have self-loops, i.e., cycles with just one edge. 
\begin{prop}[Necessary Condition]
A necessary condition for tabling to improve termination of the
computation of normal form of $t$ with respect to $P$ is the existence
of a cycle in the needs graph of $P$. 
\end{prop}
This condition can be checked in time that is linear in the sum of the
lengths of the right-hand sides of $P$ using a standard search
algorithm. We only introduce  vertices in the graph that are
necessary, i.e., they represent either defined symbols with at least
one edge or a symbol that is adjacent to a defined symbol. The reader
may wonder if it is possible to have a defined symbol with no
originating edge. This happens when all the rules defining this symbol
are  collapsing, i.e., the
right-hand side is just a variable. 

Clearly, the above condition is not sufficient for two different
kinds of reasons: (i) for instance,
$t$ could be a normal form and $P$ can have many cycles, but obviously
tabling is of no help, (ii) $P$ may have cycles but still represent a
terminating system (e.g., the fibonacci system above). Trying to
decide termination of $P$ is in general undecidable and even when it is
decidable it can be prohibitively expensive.  
Therefore, we only try to ameliorate the first source of insufficiency
by  analyzing $t$ as well.  
\begin{prop}[Necessary Condition]
A necessary condition for tabling to improve termination of the
computation of normal form of $t$ with respect to $P$ is the existence
of a cycle in the needs graph of $P$  containing a vertex that is
reachable by a directed path from a vertex representing a function
symbol or constant that appears also in $t$.  
\end{prop}
This condition can be checked by marking the vertices that appear on
some cycle in the needs graph and then running a modified search
algorithm that looks for marked vertices. 

The needs graph defined above is a refinement of the call graph from
classical compilation techniques, where procedure parameters are also
taken into account. Variations of these graphs have appeared in
other  contexts in rewriting (e.g., for narrowing
a related but different graph is used in \cite{dershosiva}),
functional and logic programming (e.g., for termination). However,
generally speaking, there are differences between such graphs in
existing literature (that we are aware of) and our approach. 
Many of the  applications here appear to be new as well. 
\subsection{When does tabling help efficiency?}
Now we focus on the second benefit of tabling, i.e., the possibility
that the computation of the normal form of $t$ would need the same
reduction step more than once. Since the tabling approach here 
shares common subexpressions to the extent possible, repeated subterms
in $t$ or its reducts do not necessarily lead to repeated reduction
steps.  
\begin{prop}[Necessary Condition]
A necessary condition for tabling to improve the efficiency of the
sequential computation of normal form of $t$ with respect to $P$ is
the existence of a node with in-degree at least one and outdegree at
least one in the needs  graph of $P$.  
\end{prop}
This condition is easily checked in time that is linear in the sum of
right-hand sides of $P$. Again, this condition is not sufficient for
two kinds of reasons. Informally, they are: (i)  $t$ has no ``need''
for this vertex, or (ii) $P$ does not ``allow'' to use this vertex from
different terms.  Again, it is quite complex to decide (ii). Hence, we
only try to ameliorate the first by analyzing $t$. 
\begin{prop}[Necessary Condition]
A necessary condition for tabling to improve the efficiency of the
sequential computation of normal form of $t$ with respect to $P$ is
the existence of a node with in-degree more than one and outdegree at
least one in the needs  graph of $P$ that is reachable via 
directed paths from two different symbols of $t$.  
\end{prop}
To check this condition we can use a standard search algorithm. 
Note that we are looking for a very ``short'' chain in the needs graph,
i.e., we are looking for the possibility that just one reduction step
can be saved. The length of the desired chain can be adjusted based on
the ratio of the time it takes to do a reduction step versus the time
it takes to lookup the results of a reduction step in the table, which
grows with the table size. 

\subsection{\bf Techniques for Efficient Reduction and Tabling}
For technical reasons, the standard operators over the predefined 
 types  are not treated as constructor functions in this section, 
 even though we do not
have any rules in the system for them. 
\subsubsection{Optimizing Reduction}
We define the following class of signatures called the 
don't-reduce  signatures that represent a subclass of normal forms
that are formed from only constructors, or constructor terms. 

\begin{defn}
(a) Base case: All constructor constants are don't-reduce signatures. 
(b) Inductive case: If classes $c_1,...,c_n$  contain only
don't-reduce signatures {\em as unreduced signatures}, 
then $f(c_1,...,c_n)$ is also a don't-reduce signature if $f$ is a
constructor.   
\end{defn}

Now note that no rule can match a don't-reduce signature.
This means that when we do leftmost-outermost reduction and we reach a class
that contains an unreduced signature which is a don't-reduce signature, then 
we need not go down (below this class) recursively to find any
matches.  Actually, if a class contains a normal form, then, of
course,  we need not explore the structure below to find any
matches. However, to detect normal forms one must match rules against
the unreduced signature of a class, whereas the detection of
don't-reduce signatures is easier since it can be done in a bottom-up
manner without any matching.  Moreover, our situation is even more
complicated with respect to detection of normal forms, since we have
built-in arithmetic and other types. It is easy to construct a
scenario where  no rules match an unreduced signature $s$ because $s$
depends on a class which contains an unevaluated arithmetic signature
as unreduced signature and once it is evaluated rules may or may not
match. Although we state this in terms of signatures here, it is clear
that this idea applies to the untabled approach as well by marking
subterms of terms that are constructor terms. 
\subsubsection{Optimizing Matching}
By analyzing the term $t$ together with the program $P$ we can
eliminate unnecessary rules, i.e. those rules that are never needed in
the computation of the normal form of $t$. For this we find all the
defined symbols of $P$ that are reachable from all the symbols in $t$ 
using a standard search algorithm on the needs graph of $P$. We then
include only rules for the reachable defined symbols in building
efficient data structures for matching, e.g., discrimination nets. In
compilation this corresponds to not generating code for the
unnecessary rules and could speed up the compilation process
itself (of course, the set of terms to be normalized must be known in
advance and specified before compilation). The generated code uses
less space.   
\subsubsection{Optimizing dependency lists in Tabling}
To identify the signatures that must be updated in the event a class
is unioned into another class, a dependency list is usually associated
with each class. It
contains all the signatures that ``depend'' on this class. Our next
optimizations are intended to reduce the size of these lists, since
processing them can get quite expensive. 

For this purpose, we  define a class of signatures called the don't-add signatures,
which is a subclass of the class of don't-reduce signatures. 
\begin{defn}
(a)  Base case: All constructor constants are don't-add signatures.
(b) Inductive case: If classes $c_1,...,c_n$  contain only
don't-add signatures, then $f(c_1,...,c_n)$ is also a don't-add
signature if $f$ is a constructor. 
\end{defn}
Note the difference in case (b) between don't-add vs. don't-reduce signatures.
All constants are, of course, never added to any dependency
list since they do not depend any class. We need (a) for the inductive
step. 

Now we have the following rule for don't-add signatures: do not add a 
don't-add signature to dependency lists. However, there is a practical
difficulty in  implementing this rule, since it is quite possible that 
$f(c_1,..., c_n)$ is a don't-add signature, but because of the insertion
of a signature into class $c_i$, class $c_i$ contains a signature
which is an ``add'' signature and hence $f(c_1,...,c_n)$ is also now an
``add'' signature. It may be possible to do this efficiently, but it
will require time and also the programming may be somewhat complex. 

Therefore, we identify a subclass of don't-add signatures, called
never-add signatures. 
\subsubsection{Never-add signatures}
In the following we assume that the rewrite system is also
non-collapsing, i.e., does not have a rule whose right-hand side is a
variable. 

While we are parsing the rules, we can 
also look for: (i) if any constructor constant is the right-hand side
of some rule (note "is" the rhs and not "is in" the rhs)
and (ii) if any standard operators over the predefined types  appear in the
right-hand side of some rule. 

Now, we define the class of never-add signatures.

\begin{defn}
(a) Base case: Every user-defined constructor constant that does not
appear as the rhs of any rule is a never-add constant.  If $t$ is  
a constant from one of predefined  types in (ii) above,   then $t$ is a
never-add constant only if 
there is no rule containing an operator that takes an
argument of the type of $t$ in the right-hand-side and the term to be
normalised also does not contain any operator that takes an argument
of type t. (b) Inductive case: If classes $c_1, \ldots, c_n$ contain only
never-add signatures and $f$ is a constructor that does not
appear as the outermost symbol of the right-hand side of any rule,
then  $f(c_1,\ldots,c_n)$ is a never-add signature. 
\end{defn}

Never-add signatures are a proper subclass of don't-add signatures,
which are easily identified in a bottom-up manner and they can never
change on the union of two classes or on the insertion of a signature
in a class containing a never-add signature. So in \smaran\, we never
add a never-add signature to the dependency list of any class. 

If the rewrite system can contain collapsing rules, then it is not
sufficient to check that no constructor constant and no
constructor function appears as the outermost symbol of any
rule. Consider, the following rewrite system for example:
\begin{eqnarray}
from(x,y) & \rarr & if(y>0,cons(x,from(x+1,y-1)),nil) \\
if(true,x,y) & \rarr & x \\
if(false,x,y) & \rarr & y
\end{eqnarray}

In this system both nil and cons do not appear as the outermost symbol
of any rule but there are terms $T$ that reduce to terms of the form nil
and cons(...). Hence, it is possible that because of a union one of
the terms from $T$ changes and so on.
In fact, it is in general undecidable whether given a term $t$ and a
rewrite system $R$ whether $t$ reduces to a constructor term. 

\section{Implementation in $LR^2$}
Some of the techniques discussed above have been implemented in $LR^2$
and have proven their effectiveness. The techniques that have been
implemented are  the identification
of don't-reduce and never-add signatures. We are currently in the
process of implementing the needs-graph analysis to help the user with
choosing the tabling or the non-tabling option in $LR^2$. 
\section{Discussion and Future Work}
In this paper, we have presented some easily computed static analysis
techniques to enhance the efficiency of tabled and untabled
computations with equational logic programs. These techniques have
implications for interpreters (and  sometimes compilers) and some are 
applicable in a wider context, e.g., functional or logic
programming. The analysis described here is a first approximation since it
ignores the arguments of functions symbols. A useful direction for
future research is to use constraint-based analysis with the needs
graph to take into account the arguments as well. Care must be taken
to ensure that the benefits of the analyses outweigh the costs. 
Other avenues are transformation of the program including partial
evaluation.  
\bibliographystyle{plain}
\bibliography{/auto/disk01/faculty/cosc/rmverma/latex/space}
\end{document}